\documentclass[conference]{IEEEtran}
\IEEEoverridecommandlockouts
\usepackage{cite}
\usepackage{amsmath,amssymb,amsfonts}
\usepackage{algorithmic}
\usepackage{graphicx}
\usepackage{textcomp}
\usepackage{xcolor}
\def\BibTeX{{\rm B\kern-.05em{\sc i\kern-.025em b}\kern-.08em
    T\kern-.1667em\lower.7ex\hbox{E}\kern-.125emX}}
\usepackage{hyperref}

\usepackage{pgfplots}
\usepackage[a4paper, hmargin=13mm, vmargin={25mm,30mm}, headsep=4mm]{geometry}

\begin{document}
\thispagestyle{plain}
\begin{minipage}{\textwidth}

{\Large IEEE Copyright Notice}

\medskip

© 2022 IEEE. Personal use of this material is permitted. Permission from IEEE must be
obtained for all other uses, in any current or future media, including
reprinting/republishing this material for advertising or promotional purposes, creating new
collective works, for resale or redistribution to servers or lists, or reuse of any copyrighted
component of this work in other works.

\medskip

Published in: 2022 11th Mediterranean Conference on Embedded Computing (MECO)

\medskip

DOI: 10.1109/MECO55406.2022.9797106

\end{minipage}

\newpage

\title{Potential of WebAssembly for Embedded Systems}

\author{\IEEEauthorblockN{Stefan Wallentowitz and Bastian Kersting}
\IEEEauthorblockA{\textit{Munich University of Applied Sciences} \\
Munich, Germany \\
stefan.wallentowitz@hm.edu, bkersting@hm.edu}
\and
\IEEEauthorblockN{Dan Mihai Dumitriu}
\IEEEauthorblockA{\textit{Midokura (Sony Group)} \\
Lausanne, Switzerland \\
dan@midokura.com}
}

\maketitle

\begin{abstract}
Application virtual machines provide strong isolation properties and are established in the context of software portability. Those opportunities make them interesting for scalable and secure IoT deployments.

WebAssembly is an application virtual machine with origins in web browsers, that is getting rapidly adopted in other domains. The strong and steadily growing ecosystem makes WebAssembly an interesting candidate for Embedded Systems.

This position paper discusses the usage of WebAssembly in Embedded Systems. After introducing the basic concepts of WebAssembly and existing runtime environments, we give an overview of the challenges for the efficient usage of WebAssembly in Embedded Systems. The paper concludes with a real world case study that demonstrates the viability, before giving an outlook on open issues and upcoming work.
\end{abstract}


\section{Introduction}

Embedded Systems are heterogeneous platforms that are deployed into a variety of settings, ranging from deeply embedded microcontrollers with tough resource constraints to powerful IoT devices with AI capabilities. Hardware platforms and software vary a lot and working with different devices can quickly become challenging.

Modern embedded software development is complex and suffers from this heterogeneity~\cite{woodward}. The tools for embedded software development are often fragmented and bound to the programming language used in a project. To support portable software for a variety of platforms and easily migrate and consolidate complex software stacks, virtualization is increasingly considered for embedded systems. Application virtual machines are an interesting solution to portable software. Such virtual machines run platform-independent applications as bytecode with strong separation properties, such as the Java Virtual Machine (JVM). A successful adoption of Java to a specific domain in embedded systems is JavaCard, which is a Java variant specifically tailored to the requirements of resource-constrained smart cards~\cite{javacard}.

\begin{figure}
    \centering
    \includegraphics[width=.45 \textwidth]{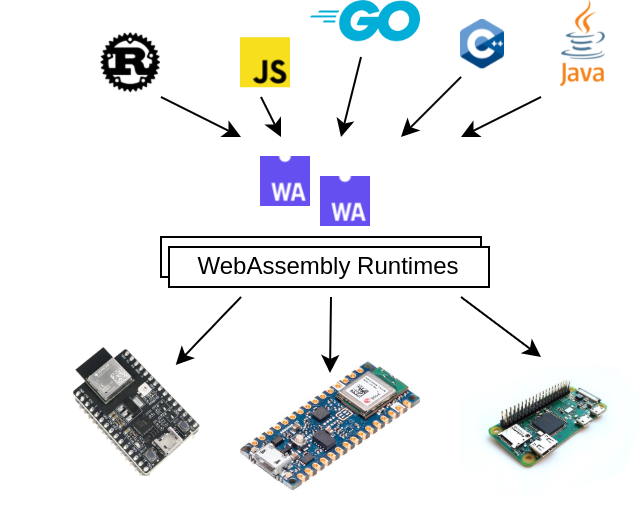}
    \caption{Portability of WebAssembly for Embedded Systems}
    \label{fig:intro}
\end{figure}

Other adoptions of Java have primarily focused on devices with user interfaces. Yet, there is still no dominance of Java for embedded systems, even despite efforts to accelerate the Java VM in hardware~\cite{jop}. One of the reasons is probably the large amount of legacy code and preferences of developers for other programming languages. While other languages could be compiled to JVM too, this potential was not adopted.

WebAssembly~\cite{wasm} is a relatively new application virtual machine format, partly comparable to JVM. Since WebAssembly was introduced in 2017 it has been quickly adopted in browsers to accelerate websites~\cite{webassemblybasics}. It has since then evolved in platforms beyond browsers in servers and desktop computers. The rich and rapidly growing, mostly open source ecosystem makes it an interesting candidate for embedded systems, too.

Figure~\ref{fig:intro} illustrates the potential of using manifold source languages to program heterogeneous hardware platforms. A WebAssembly runtime serves not only as a hardware abstraction layer (HAL), but beyond that provides strong isolation properties. WebAssembly applets can be compiled by a variety of source languages, enabled by the LLVM zoo of frontends. After porting a WebAssembly runtime to a platform, those WebAssembly applets can be executed.

In this paper we will briefly introduce WebAssembly and its fundamentals. After elaborating on WebAssembly runtimes, we will dive into the challenges ahead for making WebAssembly a suitable runtime environment for embedded systems. A case study will demonstrate the viability of WebAssembly from a real-world setup. An outlook on future directions concludes this paper.

\section{WebAssembly}

WebAssembly was initially designed with compilation target JavaScript. In early days, the Web was primarily used for exchanging documents and information in form of simple HTML pages. But with the rising complexity of Websites, JavaScript became the most (and often only) supported language for the Web. With rapid performance improvements in modern VMs, their ubiquity and widespread support, it also became a compilation target for other languages. WebAssembly is maintained by W3C~\cite{wasm}. The Bytecode Alliance~\cite{bytecodealliance} stewards runtimes and other software foundations.

WebAssembly can be compiled from a variety of source languages, ranging from script languages like JavaScript, functional languages like Elm, to low-level languages like C and C++. JavaScript’s inconsistent performance and other pitfalls that arise when compiling to it motivated the design of a new compilation target. WebAssembly “addresses the problem of safe, fast, portable low-level code on the Web”~\cite{webassemblybasics}.

WebAssembly has a lot of design goals that are also interesting for domains other than the browser. One of the problems was that the API for interacting with specific software or hardware is not properly defined yet, in contrast to the browser, where the vendors agreed on a public API~\cite{wasiannounce}. The announcement of this WebAssembly System Interface (WASI) therefore sparked a lot of interest fueled by a Twitter statement of one of the Docker inventors, Solomon Hykes: "If WASM+WASI existed in 2008, we wouldn't have needed to created Docker. That's how important it is. Webassembly on the server is the future of computing. A standardized system interface was the missing link. Let's hope WASI is up to the task!"~\cite{tweethykes}.

WebAssembly modules (also called applets) are compiled from their source language into a standardized bytecode format. A variety of (open source) compilers and frameworks are available for this. The bytecode format targets a stack-based virtual machine, basically comparable to JVM. As with other application virtual machines there are different modes to execute the bytecode. Running bytecode in an \textit{interpreter} plainly takes the applet and executes the individual commands on the stack machine. \textit{Just-In-Time (JIT)} compilation compiles the bytecode just as its executed to the target architecture on the device, which leads to significantly higher performance. \textit{Ahead-of-Time (AOT)} compilation results in a executable code for the target architecture, which seems like a reasonable compromise for embedded systems while limiting platform flexibility. Each of the approaches has their pros and cons, even interpreters can be sensible, e.g., as often found in smartcards.

\section{WebAssembly Runtime Environments}

A large number of WebAssembly runtimes has been released over the last couple of years. In the following we will compare the most prominent runtimes with a focus on popularity and validity for embedded systems during selection. Table~\ref{tab:comp_runtimes} summarizes the findings for the runtimes presented in the following.

One of the criteria for the integration into existing embedded software projects is the possibility to \emph{embed} the WebAssembly runtime. In the scenario of executing a single WebAssembly applet the following steps are executed: (i) initializing the runtime, (ii) register native functions that can be called from the applet, (iii) loading an applet from binary data, (iv) instantiating from that module , (v) call the instance's main function (until it returns), (vi) deinstantiate the module, and finally (vii) destroy the module and the runtime. Starting from those function calls more complex scenarios can be developed. The runtimes that we consider for embedded usage in the following have clear and concise interfaces, wasm3~\cite{wasm3} is for example available als an Arduino library too.

\begin{table}
    \renewcommand{\arraystretch}{1.3}
    \caption{Comparison of Popular WebAssembly runtimes.}
    \begin{tabular}{l||c|c|c|c|c}
        \textbf{Runtime} & \textbf{Language} & \textbf{AoT} & \textbf{JIT} & \textbf{Embedded}  \\
        \hline
        \hline
        wasmtime~\cite{wasmtime} & Rust & + & + & - \\
        \hline
        wasm-micro-runtime~\cite{wamr} & C & + & - & + \\
        \hline
        wasmer~\cite{wasmer} & Rust & + & + & - \\
        \hline
        wasm3~\cite{wasm3} & C & - & - & + \\
        \hline
        wasmi~\cite{wasmi} & Rust & - & - & + \\
        \hline
    \end{tabular}
    \label{tab:comp_runtimes}
\end{table}

\subsection*{wasmtime}

wasmtime~\cite{wasmtime} is a runtime maintained by Bytecode Alliance. The runtime is written in Rust but supports other languages such as C, C++ and Python. The Cranelift compiler framework is used as the compiler backend. Furthermore, the Wasmtime project drives new features of WASI, with the official specification of WASI located in the GitHub project files. In its current state it cannot be compiled for embedded targets.

\subsection*{wasm-micro-runtime}

The WASM-Micro-Runtime~\cite{wamr} is also maintained by the Bytecode Alliance and is written in C. It fully supports WASI, the latest WebAssembly features on multiple instruction set architectures like x86, Arm, XTENSA and RISC-V.  The project’s description highlights the small binary size and compiles easily for embedded platforms. It supports AOT compilation.

\subsection*{wasmer}

Another runtime that supports many features is Wasmer~\cite{wasmer}, developed by the eponymous company Wasmer. Wasmer is shipped with WASI and Emscripten support and compliant with the latest WebAssembly proposals (SIMD, Threads, etc.). Furthermore, Wasmer is configurable to support different environments. It comes with three compiler backends, each providing certain advantages, such as execution speed or compilation speed.

\subsection*{wasm3}

wasm3~\cite{wasm3} is an extremely lightweight runtime system, particularly targeted at small, resource-constrained devices. It is even available as Arduino module and can be used to interpret a wasm module on an embedded system. It does not support AoT or JIT, but could be an interesting candidate for deeply embedded devices or to run code in other languages on legacy devices that only support C.

\subsection*{wasmi}

Wasmi~\cite{wasmi} is a runtime developed by the open-source company Parity, which focuses on the blockchain and cryptocurrency implementation “polkadot”. The runtime implements an interpreter and is not capable of JiT or AoT compilation. Wasmi also doesn’t support WASI and it is unlikely that this is going to happen as the project is in “maintenance mode”, where no new features are developed. Anyhow, it compiles to baremetal Rust.

wasm-micro-runtime is the runtimes with the most supported embedded platforms and execution modes, so that we have further evaluated it for various use cases.

\begin{table}
    \renewcommand{\arraystretch}{1.3}
    \centering
    \caption{Memory Footprints of Embedded Webassembly Runtimes.}
    \begin{tabular}{l||c|c}
        \textbf{Runtime} & \textbf{Code} & \textbf{Data}  \\
        \hline
        \hline
        wasm-micro-runtime Interpreter & 94,928 & 2,068 \\
        \hline
        wasm-micro-runtime Fast Interpreter & 103,418 & 2,076 \\
        \hline
        wasm-micro-runtime AOT & 72,040 & 1,732 \\
        \hline
    \end{tabular}
    \label{tab:mem_footprints}
\end{table}

\subsection{Footprint}

Table~\ref{tab:mem_footprints} compares the footprint of wasm-micro-runtime in different execution modes on a RISC-V 32-bit platforms (ESP32 C3). As expected, the interpreter is generally larger, with performance improving methods leading to a larger code size. AOT can save around 25\% of code size of the runtime. It has to be noted that those numbers are upper limits: As modes are embedded into the target code, unused features and further code size optimizations will further reduce the sizes, along with the fact that RISC-V is considered to have a larger footprint then Arm for example.

\subsection{Benchmarking}

As shown in Table~\ref{tab:coremark} we ran the Coremark benchmark on an AllWinner V3S MCU, with 3 different configurations of wasm-micro-runtime. As expected, the wasm interpreter is quite slow. However, in AOT mode, the performance is approximately 50\% of native, which is quite acceptable for our use cases.

\begin{table}
    \renewcommand{\arraystretch}{1.3}
    \centering
    \caption{Performance Benchmark of Embedded Webassembly Runtimes on ARM32.}
    \begin{tabular}{l||c|c}
        \textbf{Wasm Runtime} & \textbf{Coremark}  \\
        \hline
        \hline
        wasm-micro-runtime Interpreter & 32  \\
        \hline
        wasm-micro-runtime AoT & 611  \\
        \hline
        Native & 1157 \\
        \hline
    \end{tabular}
    \label{tab:coremark}
\end{table}

\section{Opportunities and Challenges of WebAssembly in Embedded Systems}

As mentioned before, there are plenty of opportunities of using application virtual machines in embedded systems, in particular due to the portability and strong isolation properties. Anyhow, as the benchmarks show there are still challenges ahead. In the following we will summarize the key observations.

\subsection*{Opportunity: Portability}

Portability is one of the key arguments for virtual machines in general. While porting a container between different x86 computers with similar conditions is ubiquitous, container solutions comparable to Docker are still lacking for embedded systems. Due to platform heterogeneity, much of the code is not portable - if the same functionality is required on an IoT device with a different instruction set or operating system, it must be modified and rebuilt.

Furthermore, embedded systems code tends to be statically linked, meaning that to make any change to an application, the entire system must be rebuilt and reflashed. WebAssembly applets can instead be easily deployed, executed from flash memory and transferred between platforms. 

Beyond the target platforms there is a second portability argument to be made: Typical development on embedded systems is done in C, which excludes a significant number of developers. Allowing the (re-)use of code in different source languages with a standardized, clear compilation environment to WebAssembly can be leveraged in terms of productivity.

\subsection*{Opportunity: Isolation}

Most embedded systems do not have virtual memory. Fortunately, other memory protection methods are becoming widely adopted, but such methods are often limited and still not available in low end microcontrollers. WebAssembly and other application virtual machines are ideal for such devices, but require AOT and JIT methods to be carefully implemented.

Real Time Operating Systems are often not designed for multi-user scenarios, therefore the system calls are not well secured. Again, running 3rd party code isolated by application virtual machines helps mitigating this problem. WebAssembly and available runtimes can benefit from the experiences of firewalling in JavaCard and similar established isolation methods.

\subsection*{Challenge: Runtime performance}

As observed before, a major challenge of application virtual machines is the runtime performance of application VMs. The impact obviously depends on the execution mode of the runtime. Interpreters are often limited by their nature, so that JIT techniques are promising. AOT solves the tradeoff between runtime performance and footprint, but needs to consider underlying protection mechanisms. Isolation properties of hardware platforms and the acceleration by hardware extensions are interesting in this context.

\subsection*{Challenge: Application management}

There are many opportunities of portable applets, such as building complex applications that can be easily deployed on a variety of IoT platforms. It gives the opportunity to switch between vendors quickly. But there needs to be a level of trust before installing third party applications. Protocols and methods to load and run trusted applets need to be established. But it is not mandatory to reinvent such methods, standards such as Global Platform~\cite{gp} or similar can be adopted instead.

Furthermore, deploying large scale fleets of IoT applets onto a variety of platforms needs robust dependency management and standardized 3rd party applets that serve as hardware abstraction layers. Tools and frameworks for the management and maintenance comparable to Toit~\cite{toit} can deliver scalable, secure IoT management platforms, that significantly improve productivity.

\subsection*{Challenge: Target WebAssembly to Embedded Systems}

There are a few hurdles in WebAssembly that can be limiting for embedded systems with strict resource constraints. For example the minimum memory size of 64kB, typing of data on the stack or the absence of small integer types are often cited as limitations~\cite{specissue}. Work on revisions of the spec can address those issues, while non-standard runtime extensions are found nowadays.

\section{Case Study}

As discussed throughout this paper we believe in the future of WebAssembly as application VM in embedded systems. In the following we demonstrate the applicability to real-work use cases with a case study from the field of deep learning in IoT.

\subsection{Scenario}
We introduce a new class of IoT devices called \emph{vision based sensors}. These are meant to use visual sensing modality, but they are not for \emph{imaging} but rather for \emph{sensing} meaning that they extract metadata about the environment. Depending on the exact deployment scenario, such devices have a variety of constraints, such as power consumption, communication, and cost. Privacy may also be a constraint in some cases. These constraints dictate that most or all of the signal processing is done at the edge, on the device, so that less information is transmitted to the cloud. There may be a variety of embedded hardware platforms, each suited for a particular set of constraints. For example, some devices may need to be battery powered, in order to be deployed in the field, while others may require greater computational power, for their particular type of scenario. Furthermore, as we are working with multiple SoC (System on a Chip) vendors, there are a variety of ISAs (Instruction Set Architectures) to deal with.

\subsection{Model}
\begin{figure}
\centering
\noindent
\includegraphics[scale=0.3]{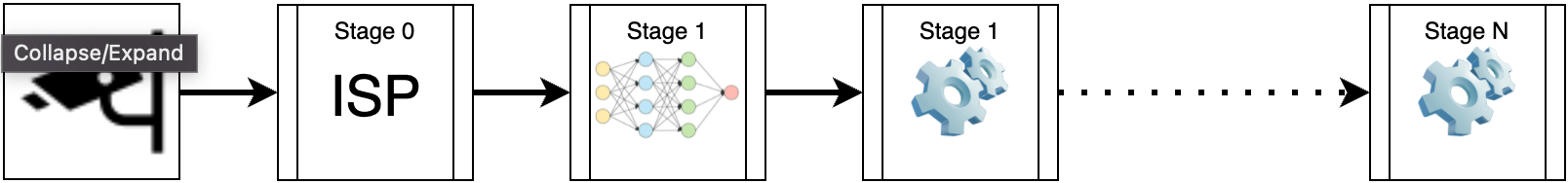}
\label{fig:pipeline}
\caption{Example of a pipeline}
\end{figure}

We model the processing as a pipeline of steps, most of which run on the application processor. See \ref{fig:pipeline}. The pipeline starts with the raw sensor input, which is then processed by an ISP (image signal processor), after which the DNN (deep neural network) inference is run, on the application processor or on a dedicated accelerator, a DSP (digital signal processor). Subsequent steps include normalizing the DNN output to a task specific representation, and then running \emph{Business Logic}. The Business Logic step requires the most flexibility, as it can be arbitrarily programmed by a developer.

This type of sensing device needs OTA (over the air) programmability, because it's function and mode of operation may be changed after deployment. For example, changing the device task from counting cats to counting dogs. Wasm gives us a safe way to change the processing pipeline, while maintaining near-native performance.

Using WebAssembly applets for the sensing pipeline stages is solves multiple problems, including isolation, runtime reconfiguration, and portability among platforms.

\section{Outlook}

WebAssembly is an interesting application virtual machine for embedded systems. While runtimes still suffer from expected performance limitations, the mostly open source work in the WebAssembly ecosystem puts a spotlight on it for consideration of containers in embedded systems.

Beyond groundwork on runtimes and their performance, we anticipate the standardization of APIs compatible to WASI in the foreseeable future. The adoption of embedded system APIs will be in the focus of our work, for example for image sensor and the ISP, or device and applet management. The interface types \verb|wit| and binding generators are interesting in this context, as they easy multi-language support and compatibility.

Finally, we believe that frameworks and tools for the efficient management of large fleets of heterogeneous systems, including IoT devices, their root-of-trust extensions, edge devices and the cloud will emerge and are actively working on such.

WebAssembly is not the silver bullet for embedded systems, but projects can already benefit when performance is not key. Runtimes focused on embedded systems can be expected to further reduce the gap to native performance. Overall, WebAssembly has the potential to have an impact on the containerized future of embedded systems.

\bibliographystyle{IEEEtran}
\bibliography{bibliography}

\end{document}